\documentclass{acm_proc_article-sp}

\usepackage{graphicx}

\title{ATLAAS-P2P: a two layer network solution for easing the resource discovery process in unstructured networks}
\numberofauthors{4} 
\author{
\alignauthor
Ranieri Baraglia\\
       \affaddr{ISTI - CNR}\\
       \email{ranieri.baraglia@isti.cnr.it}
\alignauthor
Patrizio Dazzi\\
       \affaddr{ISTI - CNR}\\
       \email{patrizio.dazzi@isti.cnr.it}
\alignauthor
Matteo Mordacchini\\
       \affaddr{ISTI - CNR}\\
       \email{matteo.mordacchini@iit.cnr.it}
       \and
\alignauthor
Laura Ricci\\
       \affaddr{University of Pisa}\\
       \email{ricci@di.unipi.it}
}

\date{ } 

\usepackage{etoolbox}
\makeatletter
\patchcmd{\maketitle}{\@copyrightspace}{}{}{}
\makeatother

\begin{document}

\maketitle

\begin{abstract}
\emph{ATLAAS-P2P is a two-layered P2P architecture for developing systems providing resource aggregation and approximated discovery in P2P networks. Such systems allow users to search the desired resources by specifying their requirements in a flexible and easy way. From the point of view of resource providers, this system makes available an effective solution supporting providers in being reached by resource requests.}
\end{abstract}

\section{System Description}
The process of discovery useful resources in a P2P network is highly conditioned to the query formulation mechanism~\cite{baraglia2010p2p}. Users should be enabled to easily express their needs and an efficient query resolution mechanism should be able both to efficiently find significant resources and to limit the number of messages exchanged. Common techniques for searching resources in P2P systems are based on range queries over a set of different attributes~\cite{icumt2009}. However, the amount of resources in a P2P network could be very large and heterogeneous, and users knowledge about the available resources could be not enough accurate to allow them to properly formulate their queries. Actually, a user is more likely able to define an ``ideal'' resource that satisfies her needs and ask to the search system to find resources close to such entity. Indeed, she would avoid the need to specify precise ranges on all the attributes. She would simply provide an example of what better suits her requirements. 

This mechanism would simplify the work for users and would lead to a more efficient exploitation of the search system. Moreover, from the point of view of resource providers, which aim is to be found by users, it would provide an effective infrastructure for resource advertising. 

ATLAAS-P2P consists in a P2P system that provides to the user a flexible way to express her requirements and an effective solution for letting resource providers be reached by users requests. It is based on a two-layer architecture, where peers in the network represent the resources of providers. The lower layer is an unstructured, gossip-based, P2P network~\cite{baraglia2012peer,baraglia2011group} allowing peers to efficiently gather in logical groups of nodes representing similar resources. The role of this layer is to automatically capture the affinities existing between resources belonging to different providers and put them in common communities. Those communities distributively elect their own representatives. The profiles of these representatives are used as the descriptors of such communities. Once elected, each representative registers itself on the higher layer, a structured, DHT-based, network.

\begin{figure}[htbp]
   \centering
   \includegraphics[width=\columnwidth]{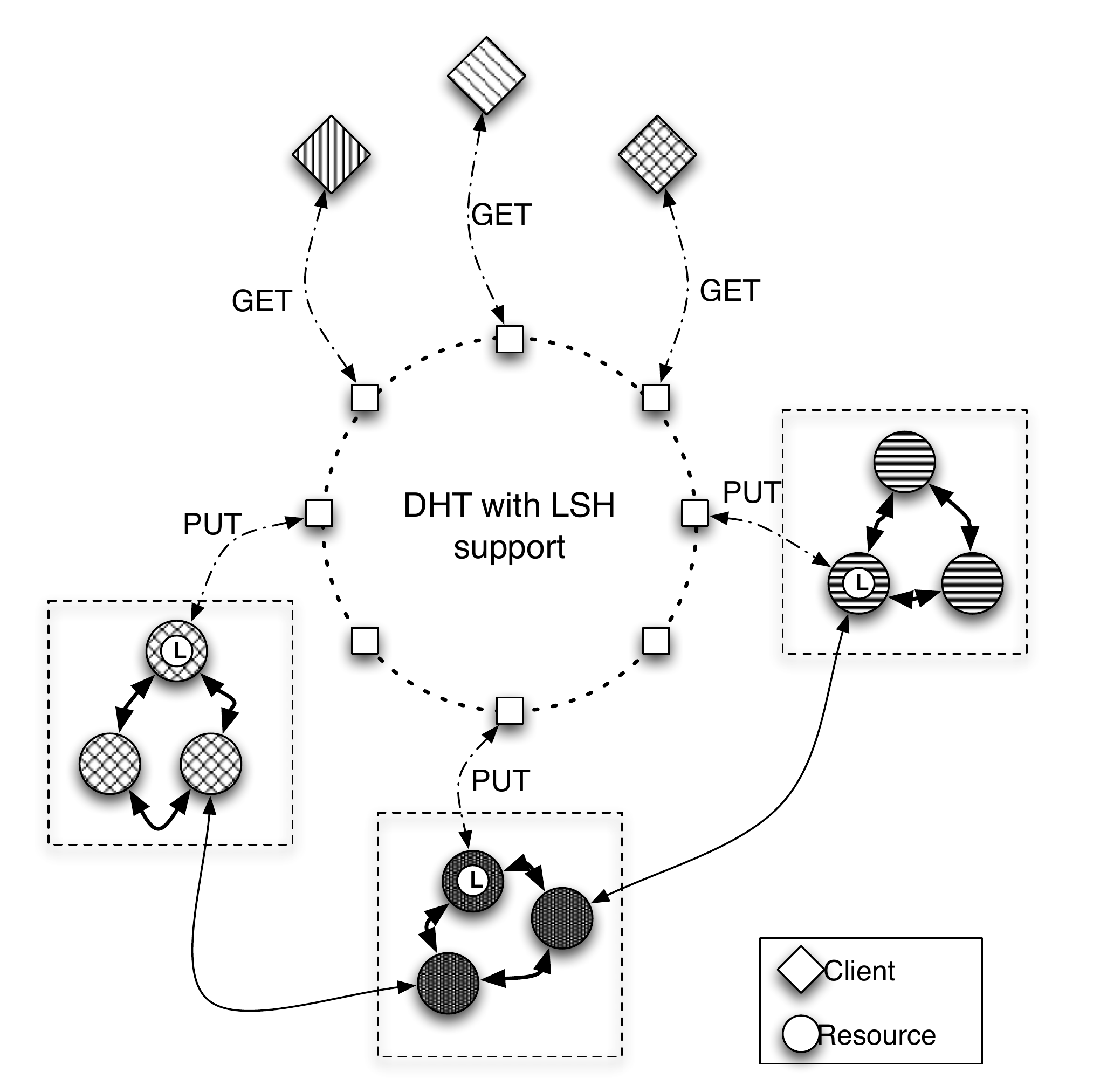} 
   \caption{Overall Architecture}
   \label{fig:overall}
\end{figure}

The structured network has been extended to support approximated searches over the community representatives~\cite{Baglini}. Users can submit to this network the queries by providing sample resources consisting in prototype of resources they are searching for. Most similar representatives are found and are selected for forwarding the query inside the community they represent in a gossip fashion. This permits to give to users suitable alternative to resources they are searching for when none of the resources available in the system matches the user request.

The overall architecture of ATLAAS-P2P is sketched in Figure~\ref{fig:overall}. Peers (circles) form distinct communities built on a similarity basis in the unstructured gossip-based layer. Each community elects a representative, denoted with an L in the figure. Each representative is in charge of register itself on the higher structured layer. Users of the systems (rhombus) can query the structured network searching for the resources they need. Results will consist of the most significative community profiles and their representatives. The representatives will act as entry points to further forward queries to the peers of the represented community. 

This architecture simplifies the task of searching for peers which profile is similar to the one given in input by a user, to the task of searching for communities described by a profile similar to the input one. This reduces both the amount of comparisons to perform and the amount of peers to contact. As a consequence, also the amount of generated network traffic decreases.

ATLASS-P2P ability to return significant resources has been tested using a dataset of 200 word domain labels organised in a hierarchical structure built by exploiting the WordNet domain~\cite{magnini:integrating}. 

\begin{figure}[htbp]
   \centering
   \includegraphics[width=\columnwidth]{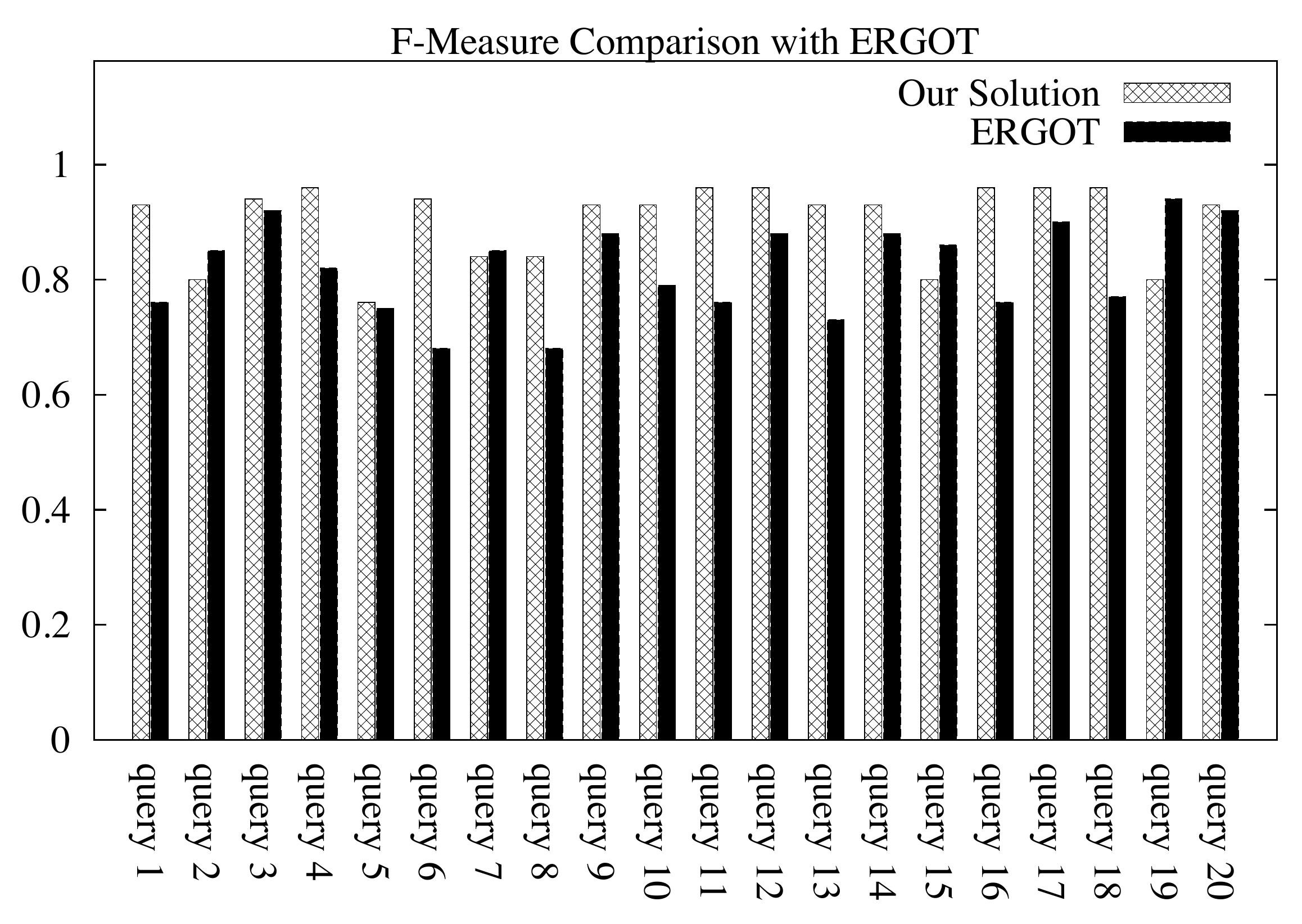} 
   \caption{Experiments}
   \label{fig:experiments}
\end{figure}

\newpage

The content of this dataset has been used for generating textual descriptions, such description have been assigned to 5000 peers according to a Zipf distribution for building the peer profiles.

ATLAAS-P2P performances are presented by Figure~\ref{fig:experiments} that compares them to the ones provided by ERGOT~\cite{Pirro:2012:DSO:2109227.2109320}, a solution aiming at the same goal but based on semantic overlay networks. 




\end{document}